\documentclass[RNAAS]{aastex62}

\begin{document}

\title{Spectroscopy of the Proposed White Dwarf Pulsar ASASSN-V J205543.90+240033.5}

\correspondingauthor{Peter Garnavich}
\email{pgarnavi@nd.edu}

\author{R. Mark Wagner}
\affiliation{Large Binocular Telescope Observatory, 933 N. Cherry Ave.
Tucson, AZ 85721, USA}
\affiliation{Department of Astronomy, Ohio State University, Columbus, OH 43210, USA}

\author{Peter Garnavich}
\affiliation{University of Notre Dame, Notre Dame, IN 46556, USA}

\author{John R. Thorstensen}
\affiliation{Department of Physics and Astronomy, Dartmouth College, Hanover, NH 03755 USA}

\author{Colin Littlefield}
\affiliation{University of Notre Dame, Notre Dame, IN 46556, USA}
\affiliation{Bay Area Environmental Research Institute, Moffett Field, CA 94035 USA}

\author{Paula Szkody}
\affiliation{Department of Astronomy, University of Washington, Seattle, WA 98195, USA}

\begin{abstract}
We obtained spectra of ASASSN-V~J205543.90+240033.5 (J2055), a system that shows photometric variations similar to the white dwarf (WD) pulsar AR~Scorpii. Our spectra display a continuum rising steeply toward the blue as well as an array of emission lines. Resolved Balmer and Paschen lines are seen with H$\alpha$ and H$\beta$ having central absorption features. The strongest lines are unresolved C~II, C~III, and N~III as well as doubly ionized helium. The spectra are similar to that of YY~Hya, and suggest that J2055 is a post-common envelope binary consisting of a hot compact star irradiating the face of a secondary of unknown spectral type. Velocity variations detected from the emission lines confirm the binary nature of J2055. The origin of the 10-minute photometric variation remains uncertain.

\end{abstract}


\section{Introduction} 

Optical photometry of ASASSN-V~J205543.90+240033.5 (J2055 hereafter) reveals periodic variations on 12-hour and 10-minute timescales \citep{kato21a,kato21b} that are similar to variability seen in the white dwarf (WD) pulsar AR~Sco \citep{marsh16,stiller18}. AR~Sco is a binary consisting of a red dwarf detached from a magnetic WD that rotates with a period of 1.95~minutes. For J2055, the 12-hour variation is thought to be related to the binary orbit while the origin of the 9.77-minute periodicity is uncertain.  

\section{Data}

We obtained spectra of J2055 using the Large Binocular Telescope (LBT\footnote{The LBT is an international collaboration among institutions in the United States, Italy and Germany. LBT Corporation partners are: The University of Arizona on behalf of the Arizona Board of Regents; Istituto Nazionale di Astrofisica, Italy; LBT Beteiligungsgesellschaft, Germany, representing the Max-Planck Society, The Leibniz Institute for Astrophysics Potsdam, and Heidelberg University; The Ohio State University, representing OSU, University of Notre Dame, University of Minnesota and University of Virginia.}) and twin Multi-Object Dual Spectrographs \citep[MODS;][]{pogge12} on 2021, September 16 (UT). A sequence of eleven, 150s exposures were taken with MODS1 through the SX telescope providing coverage over 36~minutes including readout and overheads. The same exposure sequence was obtained with MODS2 through the DX telescope. The image quality averaged 0.7 to 0.8 arcseconds during the observation. A 0.8~arcsec wide slit was employed with the gratings to give a spectral resolution in the red of 1350, or 215~km$\;$s$^{-1}$ (FWHM) at H$\alpha$. In the blue, the resolution is 1890, or 160~km$\;$s$^{-1}$ (FWHM) at H$\beta$. Each spectrograph has a red and blue arm split at 5650~\AA\ by a dichroic mirror allowing wavelength coverage from 3200~\AA\ to 1.01~$\mu$m. 

The spectra were taken near photometric phase 0.04 based on the ephemeris from \citet{kato21b}. This is very close to the peak brightness of the 12-hour photometric periodicity.

\vspace{0.5cm}
\section{Analysis}

\begin{figure}[h!]
\begin{center}
\includegraphics[scale=0.55,angle=0]{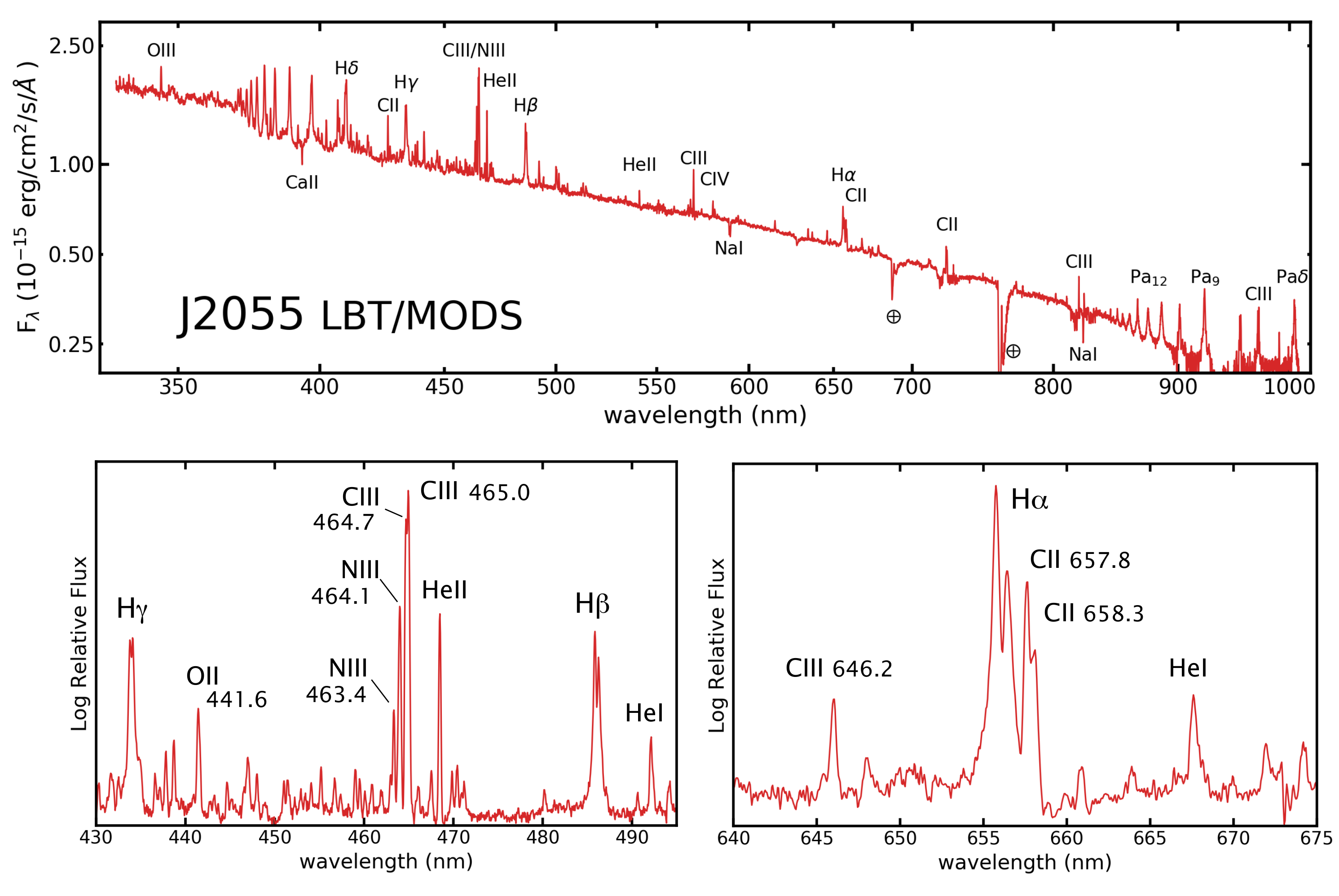}
\caption{{\bf Top:} The average spectrum of J2055 from the LBT/MODS observations. The Balmer and Paschen series are seen in emission as well as narrow carbon and nitrogen lines. Interstellar or circumbinary absorption features of Na~I and Ca~II are visible. {\bf Lower-left:} Close-up of the Bowen features of C~III/N~III and He~II. {\bf Lower-right:} Close-up around H$\alpha$ showing the deep central absorption and broad wings. The C~II blend around 658~nm is unusually strong relative to H$\alpha$.  \label{fig1}}
\end{center}
\end{figure}

The resulting average spectrum is displayed in the top panel of Figure~\ref{fig1}. It shows a continuum rising toward the blue with
a large number of narrow emission lines from the ultraviolet to the near infrared. The Balmer and Paschen hydrogen emission sequences are evident.
However, the strongest lines are permitted carbon and nitrogen emission features with a range of ionization states. For example, the blend of C~III lines at 4649~\AA\ is 50\%\ brighter than H$\beta$. Also, the C~II 6578+6583~\AA\ is half the flux of H$\alpha$. He~II emission is also significant at 4686, 5411, and 8237~\AA .

H$\alpha$ and H$\beta$ show a central absorption that vanishes toward the higher Balmer lines. The H$\alpha$ and H$\beta$ line widths are 600~km$\;$s$^{-1}$ (FWHM). The base of the H$\beta$ line extends to $\pm$900~km$\;$s$^{-1}$.  The Balmer widths decrease toward the higher level transitions with H$\epsilon$ having a width of 440~km$\;$s$^{-1}$ (FWHM). The Balmer decrement is inverted, with the H$\alpha$/H$\beta$=0.5. The He~II and carbon lines are unresolved at this resolution.

Cross-correlation between the average spectrum and each of the 11 exposures shows a systematic decrease in velocity totalling 44~km$\;$s$^{-1}$ over the 36 minutes of data. This appears to confirm the binary nature of the system, although the spectra cover only 5\%\ of the photometric period. Narrow absorption features are detected from Na~I and Ca~II. The absorption lines do not show any significant variation in velocity suggesting they originate in circumstellar or interstellar gas.

\section{Conclusions}

The spectrum of J2055 near maximum is similar to the post-common envelope binaries (aka pre-cataclysmic binaries) TW~CrV \citep{tw_crv} and YY~Hya \citep{yy_hya}. These systems consist of a hot compact star, probably a WD, irradiating the face of a late-type (K or M type) star. No evidence of the cool secondary is seen in our spectra, possibly because they were obtained near maximum light when the irradiated face dominates the flux.

The optical spectrum of J2055 is clearly quite distinct from that of AR~Sco \citep{marsh16,garnavich19}. The WD in AR~Sco has a relatively cool surface temperature \citep{garnavich21}, leading to a spectrum dominated by Balmer, neutral helium, and Ca~II emission lines. The heating of its secondary may be enhanced by non-thermal radiation. The temperature of the WD in J2055 appears much higher than for AR~Sco, and post-common envelope systems often show enhancement of CNO elements \citep{shimansky11}. The differences in spectral properties between J2055 and AR~Sco are likely due to the combination of enhanced CNO abundances and higher temperatures in J2055.

The origin of the 9.77-minute periodicity \citep{kato21a} remains uncertain. It may come from surface temperature variations on a rotating WD, or, as in AR~Sco, the beat period arising from the interaction of the WD magnetic field interacting with the secondary star.

\acknowledgments

We thank S. Kimeswenger and co-authors for providing a preview of their paper. We thank Jenny Power, Olga Kuhn, and LBT staff for their assistance in obtaining the MODS spectra of J2055 during the observatory restart block.


\begin{thebibliography}{}

\bibitem[Garnavich et al.(2019)]{garnavich19} Garnavich, P., Littlefield, C., Kafka, S., et al.\ 2019, \apj, 872, 67. doi:10.3847/1538-4357/aafb2c

\bibitem[Garnavich et al.(2021)]{garnavich21} Garnavich, P., Littlefield, C., Lyutikov, M., et al.\ 2021, \apj, 908, 195. doi:10.3847/1538-4357/abd4db

\bibitem[Kato(2021)]{kato21a} Kato, T.\ 2021, arXiv:2108.09060

\bibitem[Kato et al.(2021)]{kato21b} Kato, T., Hambsch, F.-J., Pavlenko, E.~P., et al.\ 2021, arXiv:2109.03979

\bibitem[Kimeswenger et al.(2021)]{yy_hya} Kimeswenger, S., Thorstensen, J.~R., Fesen, R.~A., et al.\ 2021, arXiv:2110.03935

\bibitem[Marsh et al.(2016)]{marsh16} Marsh, T.~R., G{\"a}nsicke, B.~T., H{\"u}mmerich, S., et al.\ 2016, \nat, 537, 374. doi:10.1038/nature18620

\bibitem[Pogge et al.(2012)]{pogge12} Pogge, R.~W., Atwood, B., O'Brien, T.~P., et al.\ 2012, \procspie, 8446, 84460G

\bibitem[Shimansky et al.(2016)]{tw_crv} Shimansky, V.~V., Mitrofanova, A.~A., Borisov, N.~V., et al.\ 2016, Astrophysical Bulletin, 71, 463. doi:10.1134/S199034131604009X

\bibitem[Shimansky et al.(2011)]{shimansky11} Shimansky, V.~V., Bikmaev, I.~F., \& Shimanskaya, N.~N.\ 2011, Astrophysical Bulletin, 66, 449. doi:10.1134/S1990341311040079


\bibitem[Stiller et al.(2018)]{stiller18} Stiller, R.~A., Littlefield, C., Garnavich, P., et al.\ 2018, \aj, 156, 150. doi:10.3847/1538-3881/aad5dd

\end{thebibliography}
\end{document}